# Heat current anticorrelation effects leading to thermal conductivity reduction in nanoporous Si

Laura de Sousa Oliveira[1*], S. Aria Hosseini[2], Alex Greaney[2], Neophytos Neophytou[3]

[1]Chemistry, University of Wyoming, Laramie, WY 82071, USA

[2]Mechanical Engineering, University of California, Riverside, CA 92521, USA

[3]School of Engineering, University of Warwick, Coventry, CV4 7AL, UK

[*] Laura.deSousaOliveira@uwyo.edu

## Abstract

Prevailing nanostructuring strategies focus on increasing phonon scattering and reducing the mean-free-path of phonons across the spectrum. In nanoporous Si materials, for example, boundary scattering reduces thermal conductivity drastically. In this work, we identify an unusual anticorrelated specular phonon scattering effect which can result in additional reductions in thermal conductivity of up to ~ 80% for specific nanoporous geometries. We further find evidence that this effect has its origin in heat trapping between large pores with narrow necks. As the heat becomes trapped between the pores, phonons undergo multiple specular reflections such that their contribution to the thermal conductivity is partly undone. We find this effect to be wave-vector dependent at low temperatures. We use large-scale molecular dynamics simulations, wave packet analysis, as well as an analytical model to illustrate the *anticorrelation effect*, evaluate its impact on thermal conductivity, and detail how it can be controlled to manipulate phonon transport in nanoporous materials.

I. INTRODUCTION

Nanostructuring has enabled an unprecedented control of phonon transport with widespread applications ranging from microelectronic devices [1] to data storage [2], and micro-electromechanical systems [3,4]. Strong focus has been placed on reducing thermal conductivity for thermoelectric and heat insulation applications [5-8]. This is largely because nanostructuring can significantly reduce a materials' thermal conductivity — particularly in semiconductors and dielectrics. Reductions of up to two orders of magnitude have been observed over the last few years in several Si-based nanostructures, including rough Si nanowires [9,10], thin films [11,12], and Si-based alloys and superlattices [13]. Recent works concerning Si-based nanoporous materials, have also shown that the room temperature thermal conductivity can be reduced beyond the materials' amorphous limit [5-7]. The consensus is that porosity reduces the heat capacity of the material and thus its thermal conductivity to some degree, but the additional scattering of phonons on the pore surfaces reduces the thermal conductivity even below the amorphous limit.

Existing nanostructuring strategies largely focus on alloying and introducing defects at different length scales to increase 'incoherent' phonon scattering and thereby reduce the relaxation times of phonons across the spectrum. Most often, the phonon-gas picture suffices to describe thermal conductivity, even in nanoporous materials [14-16]. However, the issue as to whether 'coherent' wave effects alter the phonon dispersion relations — changing group velocities, the density of states, and creating phononic bandgaps — or result in the localization of modes is still an open topic [6,16-19]. Herein, we report on the emergence of anticorrelated (AC) specular phonon scattering (and thus heat flux) as a result of heat trapping between the pores, which can provide up to ~ 80% *additional* reduction in thermal conductivity for specific nanoporous geometries. Anticorrelated heat flux has been observed in amorphous and fluid materials, but not in crystalline materials [20-23]. The AC effect can benefit thermoelectric applications, but also find wider application for the control and manipulation of heat carrying phonons in nanophononic metamaterials in general. In this work we use large-scale equilibrium molecular dynamics (EMD), wave packet simulations, and develop an analytical model to: (1) describe the anticorrelated heat flux behavior and the conditions that preclude it (heat trapping), (2) evaluate its impact on thermal conductivity, and (3) detail how this effect can be controlled to manipulate phonon transport in nanoporous materials. We begin by describing the approaches used (Section II), followed by the observations of anticorrelated heat flux behavior as a function of the nanoporous geometries (Section III), obtained with the Green–Kubo approach. In Section IV, we discuss the results of wave packet simulations, which provide an illustrative picture of the underlying physical mechanism for the heat flux anticorrelation. These results indicate that heat trapping and multiple reflections between the pores is what manifests as anticorrelation effects in the heat flux and further suggest the effect to be wave-vector dependent. Finally (in Section V), a simple ray tracing model is introduced, which connects the behavior observed in Sections III and IV by showing, in a simple manner, how specular phonon reflections between the pores with multiple reflections can lead to heat flux anticorrelation effects. Section VI offers a conclusion for the bulk of the work presented.

## II. METHODS

Thermal transport in nanostructured geometries requires an understanding beyond what is achievable at a continuum level, and yet simulation domains larger than can be accommodated by first principles approaches. Methodologies available to study nanoporous morphologies are thus limited to classical molecular dynamics (MD) [24-27], or semi-classical approaches involving the numerical solution of the Boltzmann transport equation (BTE) [28-32], and, to some extent, lattice dynamics [24,33,34]. In this work, we use a combination of both equilibrium molecular dynamics (EMD) and wave packet simulations to evaluate thermal transport in Si nanoporous structures. We further develop a simple statistical model of a grey population of heat-carrying acoustic phonons to illustrate how heat trapped between the pores can lead to anticorrelated behavior in the HCACF. The model is described in Section V.

The Green–Kubo is a well-established approach to determine the thermal conductivity of a system from its thermal fluctuations at equilibrium, such that the thermal conductivity along $x$ (i.e., the length of the simulation cell as shown in the inset in Fig. 1 (a)), $\kappa_x$, is given by:

$$\kappa_x = \frac{V}{k_B T^2} \int_0^\infty \langle J_x(t) J_x(t+\tau) \rangle \, d\tau, \quad (1)$$

where $V$ and $T$ are the volume and temperature of the system respectively, $k_B$ is Boltzmann's constant, and $\langle J_x(t) J_x(t+\tau) \rangle = A(\tau)$ is the averaged but non-normalized heat current autocorrelation function (HCACF) of the $x$-component of the instantaneous heat-flux, $J_x(t)$, at simulation time $t$. The HCACF measures the size and longevity of thermal fluctuations in the heat flux of a system in equilibrium, and is central to revealing the anticorrelated behavior that we describe below.

Simulations were performed with the large-scale molecular dynamics software LAMMPS [35], using the Stillinger–Weber (SW) potential [36]. We have opted to use the Stillinger–Webber potential in this work in part because it is commonly used to model heat transfer in silicon [24,37], and for consistency with our previous work [14]. Although it overestimates the thermal conductivity [38], the Stillinger–Webber potential provides a reasonable match for the phonon dispersion relations, in particular for the acoustic phonons [39]. The results were averaged for sets of 15–20 simulations to mitigate the large uncertainty in the Green–Kubo approach [14], and smaller simulation cell sizes were used where possible to reduce computational expense. Simulation cell sizes varied between 40×10×10 and 200×10×10 unit cells for the simulations associated with Figs. 1 and 3, and 100×10×10, and 100×24×10 unit cells for the simulations in Fig. 2. This corresponds to dimensions ~ 21.72×5.43×5.43 to ~ 108.6×5.43×5.43 nm$^3$ for Figs. 1 and 3, and ~ 108.6×5.43× 5.43 to ~ 108.6×13.03× 5.43 nm$^3$ for Fig. 2, respectively. System sizes are indicated in the figures, and following common practice we report the fractional change in thermal conductivity compared to the pristine system, $\kappa_{porous}/\kappa_0$. We considered porous Si as illustrated in the inset of Fig. 1 (a), where the pores are empty cylindrical regions 'etched' from the top all the way to the bottom of the material. The systems were brought to and equilibrated at ~300 K, such that each system has its own initial configuration. Temperature equilibration is done in two parts: (1) the systems are brought to room temperature and allowed to thermally expand in the isothermal, isobaric ensemble (NPT) for 125 ps, and (2) equilibrated in the microcanonical ensemble (NVE) for an additional 125 ps, before any calculations are performed, also in NVE, for 10 ns. Equilibration is performed using a 0.5 fs interval, whereas a 2 fs time step is used to record the heat flux for the HCACF calculation. Transport properties reported in this work are computed along the *x*-axis, that is, along the long direction of each simulation cell (as shown in the inset in Fig. 1 (a)), which is aligned with the [1 0 0] crystal direction. Additional simulation details can be found in our recent work [14].

For illustration of phonon propagation in the structures we simulate, we form Gaussian phonon wave packets and let them propagate while we monitor their trajectory. A Gaussian phonon wave packet is a propagating wavefunction formed by a linear superposition of plane waves weighted by a Gaussian distribution around a localized wavevector, and defined by

$$u_{lj\mu\gamma} = \sum_{q}^{q_{BZ}} \mathcal{A}_o \left(\frac{1}{\sigma\sqrt{2\pi}}\right) e^{\left(\frac{q-q_o}{\sigma\sqrt{2}}\right)^2} \epsilon_{j\mu} e^{-i(r_l q + \omega_\gamma t)}. \qquad (2)$$

Here, $u_{lj\mu\gamma}$ is the displacement of the j$^{th}$ atom in the l$^{th}$ unit cell along a direction μ (in *x*, *y* or *z*) for a given mode, γ. $\mathcal{A}_o$ is the amplitude of the wave packet, which can be tuned to a desired wave packet energy. The wave packet is centered at a desired carrier wave vector, $q_o$, with an uncertainty in momentum space specified by σ. The term $\omega_\gamma$ is the frequency of the mode γ at $q_o$, and $\epsilon_{j\mu}$ is the eigenvector of the j$^{th}$ atom along μ at the selected mode, γ. $r_l$ is a vector that points to the l$^{th}$ unit lattice, and t is the time. The sum over q is performed for all wave vectors in the first Brillouin zone that are commensurate with the compute cell. The phonon wave packet simulations are centered at wave vectors $q_o$ = 0.46, 0.93, 1.45, 1.74, 2.89, 4.05, and 5.21 nm$^{-1}$, along the [100] crystal direction for both the longitudinal and transverse acoustic modes (γ). Large simulation cells, consisting of 800×10×10 unit cells for Si are used to precisely model the wave packets with very fine uncertainty in momentum space (σ = 0.05 nm$^{-1}$). The initial position of the atoms in the MD simulation is computed using Eqn. (2) and the initial velocity is computed from the derivative of $u_{lj\mu\gamma}$ with respect to time ($v_{lj\mu\gamma} = \frac{du_{lj\mu\gamma}}{dt}$).

In our wave packet simulations the system is initially at 0 K and the wave packets are added with $\mathcal{A}_o$ for each wave packet tuned so that it raises the temperature of the system by around ~5 K, rather than adding one phonon, $\hbar\omega$, of energy. The reason for this is that, in the system sizes modeled, $\omega\hbar$ for wave packets near the Brillouin zone center is too small to be resolved above the numerical noise, while a single $\hbar\omega$ would raise the system temperature by hundreds of Kelvin for wave packets near the Brillouin zone edge. While the Green–Kubo calculations were performed at 300 K, performing the wave packet simulations at lower energy (~5 K) help keep phonon thermalization at bay, such that the acoustic frequencies selected for the packets don't easily decay into other modes/frequencies due to anharmonicity. This allows us to observe the scattering behavior of specific wavevector phonons at the nanopores, as they are less likely to be obfuscated by anharmonic effects. The actual values of $\mathcal{A}_o$ are included in the SI [40].

## III. THE ANTICORRELATION EFFECT

In this section, we clarify and demonstrate the emergence of anticorrelations [41] in the heat flux and quantify its effect on the thermal conductivity as a function of pore and neck sizes, as well as pore periodicity. Three sets of geometries are considered in Fig. 1: (1) pores with a 1 nm radius (blue, green, orange), (2) pores with a 1.5 nm radius (red, purple, and maroon), and (3) an elongated pore (cyan), such that the 'neck', i.e., the distance between the edges of the pores perpendicular to the direction of transport (i.e., in the $y$-direction), is 1 nm, but its porosity is the same as that of the geometry in purple. A cross-section along the $xy$-plane is shown in Fig. 1 (c) for each of the geometries. The length of the simulation cell varies between ~ 27.2 and 108.6 nm, and the width along the $y$ direction is ~ 5.43 nm for all of the geometries shown in Fig. 1. The moving average of the HCACFs and the HCACF cumulative integrals for these geometries are plotted in Figs. 1 (a) and 1 (b), respectively.

The HCACF can be decomposed into contributions from short and long length scale interactions by fitting the HCACF to a sum of exponentials [20], from which the relaxation times for different length-scale phonon processes can be extracted. The rate of decay of the HCACF is thus a measure of the relaxation times of the heat-carrying phonons in the system. The typical HCACF for Si decays exponentially and monotonically to zero. In our simulations, the geometries with pore radius r = 1 nm (first triad of Fig. 1 (c) structures), and the elongated pore geometry (cyan), all match this behavior; the small oscillations around zero stem from statistical noise in the HCACF and are to be expected [14,42]. However, the HCACFs of geometries containing uniformly distributed 1.5 nm pores show *anomalous* behavior (second triad of Fig. 1 (c)). The HCACFs become negative and decay to zero from below the $x$-axis. The negative correlation or, equivalently, the a*nticorrelation* (shown in Fig. 1 (a) as the region of the HCACF that is below zero) in the narrow neck systems occurs when heat flux fluctuations in one direction are followed by fluctuations in the opposite direction. This anticorrelation reduces the system's thermal conductivity, creating a peak in the cumulative HCACF integral at the point where the HCACF crosses the $x$-axis as shown in Fig. 1 (b). We can observe in Fig. 1 (a) that there are variations in the width, the minimum, and the time after which each minimum occurs for the geometries shown. How these characteristics are affected by the geometry is discussed later in the text. We begin by quantifying the reduction in thermal conductivity that is due to the anticorrelation of the heat flux.

As a metric of the reduction in the thermal conductivity, $\kappa$, due to the anticorrelation (AC) we consider the height of the peak in the cumulative HCACF integral above its final converged value. This value is indicated in Fig. 1 (b) for the geometry in purple, and corresponds to a 21.5±6.5% decrease in the accumulated $\kappa$. For a more realistic comparison of the decrease in thermal conductivity due to this effect, we considered a reference geometry (in cyan) without AC effects (its pores are elongated, such that the neck matches those of the geometries in the first triad (top) of Fig. 1 (c)), but with equivalent porosity and number of scatterers as the geometry in purple (which belongs to the second triad (bottom) of Fig. 1 (c)). Comparing the two geometries, the purple system yields a 37.4±9.0% decrease in thermal conductivity (see inset in Fig. 1 (a)).

This estimated 37.4±9.0% reduction in thermal conductivity suggests that using the HCACF peak height to estimate the effect of AC, which yields an estimated 21.5±6.5% change in thermal conductivity (see Fig. 1 (b)) underestimates the total reduction in $\kappa$. However, it would be too computationally expensive to compute similar elongated pore geometries to match each of the other geometries investigated herein, and henceforth we use the peak height to compare the impact of each geometry on the thermal conductivity. Overall, the AC effect provides an additional path to reduce the thermal conductivity, and the scale of the reduction is comparable to that achieved by increasing the number and surface area of the scattering features. For instance, the thermal conductivity of the geometry in red (with AC effects) has a similar porosity to the geometry in green (without AC effects), which has a higher number of scatterers and thus also surface area (see inset in Fig. 1 (a) and Fig. 1 (c)). The same equivalence can be observed between the geometry in purple and the system in orange.

To further determine how the AC effect emerges with respect to the porous geometry, in Fig. 2 we examine multiple geometries with different pore and neck sizes. To vary the pore and neck sizes independently, we selected four equal length sets of geometries, but with varying width (i.e., the *y*-direction in the simulation cell), as illustrated in the insets in Figs. 2 (a) and 2 (b). This allows us to consider geometries with the same neck size, but different pore size, and vice-versa, and in general various neck and pore sizes. Multiple pore sizes are considered for each set of the characteristic geometries of different widths (blue, red, green, and cyan as shown in the insets of Fig. 2 (a)). Pore sizes are indicated in the caption of Fig. 2. In all cases the distance between the pores' center is kept constant. The length of the simulation cell is also fixed at 54.3 nm. Fig. 2 shows the HCACFs (Fig. 2 (a)), and corresponding cumulative integrals (Fig. 2 (b)) for the various geometries. Much larger dips in the HCACF can be engineered compared to Fig. 1, some affecting the cumulative integrals and the thermal conductivity in a drastic way.

If we again consider the height of the peak created in the cumulative integral of the HCACF (Fig. 2 (b)) to estimate the effect of the AC in the heat flux, as we have done before, we find that for any given geometry type (blue, red, green and cyan geometries), the percentage change in $\kappa$ plateaus as a function of the ratio between the pore radius and the neck (Fig. 2 (c)). The 'neck' (see inset illustration in Fig. 2 (b)) plays a major role in producing an anticorrelation in the heat flux. Fig. 2 (c) illustrates that (1) the AC effect is better correlated to the ratio of the neck to pore radius than to either the pore radius, or neck width alone (see inset in Fig. 2 (c)). In other words, a smaller pore system requires a smaller neck to yield the same dip minima as a larger pore system, and the ratio between the neck and pore diameter is a better metric of the total thermal conductivity than either the neck or pore diameter alone (see Fig. S5, in the SI [40]). This is consistent with results from Monte Carlo simulations as well [43]. Similarly, we intuitively expect a smaller neck and a larger radius to reflect (back) phonons more effectively and create a stronger anticorrelation effect. The results in Fig. 2 (c) further indicate that beyond a certain point it makes no difference if the neck is further reduced for a given pore size. One possibility for this plateau, which is reached at a 79.3±12.6% reduction in $\kappa$ for a fixed pore periodicity of 54.3 nm (see Fig. 2 (c)), is that a limit is reached at which point the remaining phonons, with short mean-free-paths (smaller than the pore periodicity), will thermalize before anticorrelated behavior can take place.

It follows from Fig. 2 that the extent to which the heat flux is anticorrelated can be controlled by carefully selecting the pore and neck sizes, that is, the lateral spacing between pores. Next, we show that the spacing between the pores in the direction of transport can also affect the location of the HCACF dip minima. We consider a set of geometries of periodically arranged pores, depicted in Fig. 3 (e), with the same pore radius (r = 1.5 nm) but varying pore density. All of the geometries in Fig. 3 (e) exhibit anticorrelated heat flux behavior to some extent, with the exception of the higher pore concentration geometry (in green). Each of the dips in the (moving average of the) HCACFs shows up at a different instance of the HCACF time, as seen in Fig. 3 (a). Similarly, the HCACF cumulative integral (Fig. 3 (b)) shows peaks (labelled with stars) moving to the left. The percentage change in thermal conductivity due to the AC effect is indicated in Figs. 3 (b) and (c). Notice that the peaks occur earlier than the dip minima (labelled with dots), as they correspond to the instance where the HCACF becomes negative. Interestingly, we find that there is a linear correlation between the distance between the pores, $d$ (see Fig. 3 (e)), and when the anticorrelation dip minima and integral peak maxima occur (Fig. 3 (d)). The slopes of the aforementioned relationships are also indicated, in units of velocity.

The dip minima indicate the simulation time intervals at which the anticorrelation effect is strongest. Using this measure, we find the slope of the correlation between when the anticorrelation effect is strongest (i.e., the dip minima) and $d$ to be $v_{dips}$ = 2845 m/s. (If instead we use the peaks of the cumulative HCACF, we obtain $v_{peaks}$ = 4395 m/s.) As a reference to the reader, near the Γ point of the Si phonon spectrum obtained with the Stillinger–Webber potential, the velocities for the (dominant) longitudinal acoustic and transverse modes are ~ 8100 m/s and ~ 5000 m/s respectively, yielding an average speed of ~ 6033 m/s for the three modes. There is clearly a linear relation between the appearance of the anticorrelation effects and the pore distances.

III. THE HEAT TRAPPING ORIGIN OF THE ANTICORRELATION EFFECTS

To investigate the origin of the AC effect on the porous structures, wave packets centered at a wave vector $q_o$, as detailed in Section II, are propagated through two sets of systems with pore radii of 1 nm (left-hand plots of Figs. 4 (a)–(h)) and 2 nm (right-hand plots of Figs. 4 (a)–(h)) with corresponding 3.4 and 1.4 nm necks. When we evaluate the thermal conductivity of the structures in the left and right columns with the Green–Kubo approach, the structure in the left column does not show AC effects, whereas the one in the right does. Both longitudinal and transverse modes are considered for several values of $q_o$, for which heat maps are obtained showing the evolution of the kinetic energy of the wave packets in the geometries during simulation time. The values of $q_o$ are noted in each sub-figure, and they are also indicated by the vertical lines in Fig. 4(k) on the q-axis, where the frequencies and velocities of the modes are plotted as well. This is shown in Figs. 4 (a)–(h) for the transverse mode with polarization perpendicular to the pore height (labeled TA⊥ in Fig. 4 (i)). The longitudinal (LA) and parallel transverse (TA∥) modes, as well as other $q_o$-centered packets for the same (TA⊥) mode are shown in the SI [40]. In the heatmaps (Figs. 4 (a)–(h), (j) and (l)) the ordinate indicates the propagating time and the abscissa the length direction. The positions of the pores are indicated by the white vertical lines and are located at 216 and 270 nm.

From Figs. 4 (a)–(h) it is evident that (1) the amount of heat reflected at the first pore is consistently greater for the larger pore, narrower neck geometries (center column in Fig. 4), and (2) the amount of heat that is transmitted after the second pore is significantly reduced for the same narrow neck geometries. As a consequence of hindering heat propagation through the spacing between the pores, it can also be seen that for the narrower neck geometry (center column of Fig. 4) phonons become trapped between the pores, causing the packets to oscillate back and forth. This effect is also clearly shown to be q-dependent, in that it is more or less prevalent at different wave vectors; roughly, the heat blocking and accumulation becomes stronger for phonons with larger wavevectors. This is most noticeable by considering the transmission at the second pore (i.e., the amount of kinetic energy that reaches past the second pore, located at 270 nm), which disappears for $q_o$ values of 4.07 nm$^{-1}$ and 5.24 nm$^{-1}$, but is present at other values of $q_o$.

We remark that, in the case of the wave packets, the pore neck/size controls the amount of energy trapped between the pores not only by not allowing heat to escape once through the first pore, but also by limiting the amount of heat that goes through the first pore. For instance, for $q_o$ = 4.07 nm$^{-1}$ in the narrow neck structure (Fig. 4 (g)) the intensity of the heat bouncing back and forth is somewhat less than that in the wider neck case (Fig 4 (c)), however, this is most likely because most of the kinetic energy is reflected at the first pore, and less energy is therefore available to be reflected between the pores. The multiple reflections observed in the geometries with narrower neck/larger pores corroborate the EMD simulation results discussed in the previous section, which exhibit HCACFs with negative values evidencing an anticorrelation of the heat flux. In short, like the packets which bounce back and forth between the pores for geometries with narrower neck, in the equilibrium calculations, heat similarly fluctuates back and forth between the pores. As the necks become smaller and the pores larger, more heat is trapped and scattering between the pores is intensified. This agrees with the observations in Fig. 2, which show an increase in the (proportional) amount of anticorrelated heat flux for narrower neck geometries (over positively correlated heat flux). In Fig. 2 (a), this is evident in how negative the HCACF becomes as the pore sizes increase and necks decrease.

The transmission through the first pore as a function of $q_o$ is shown in Fig. 4 (i) (see the SI [40] for calculation details) for all acoustic modes. Overall, the transmission is lower for larger wave vectors (and thus also higher frequencies). This could account for why $v_{dips}$ (Fig. 3 (d)) is *less than* exactly half the speed of the average of the acoustic modes. In other words, if higher $q_o$ phonons are most noticeably trapped between the pores, the overall velocity of these modes would be lower than their velocity at Γ, because the velocity of the modes decreases a function of q (see Fig. 4 (k)). In fact, higher frequency modes are more likely to be scattered between the pores, while larger wave vectors are less likely to 'see' small size features [43]. That said, the increased scattering rate between pores for larger pores/smaller necks is nevertheless still present at low wave vectors (see Fig. 4 (a)). Due to their weaker Umklapp scattering, phonons with small wave vectors are known to contribute more to the overall thermal conductivity than large wave vector phonons. On the other hand, it is also known that the contribution of higher q values becomes more noticeable once lower q-values have been scattered, for instance due to defect scattering [44-47].

The trend observed in Fig. 3 (a), whereby the HCACF dip moves left as the periodicity between the pores becomes smaller (i.e., as the pores become closer together in the direction of thermal transport) can be understood as a function of the maximum possible correlation distance (and thus time) for heat scattering between the pores in each case. Consider the heat trapped between the pores in the packet simulations: the distance between the pores dictates the maximum distance heat may travel from the moment it first crosses the left-hand-side pore until it finally dissipates. For this reason, the maximum correlation interval, $\tau$, is smaller when the pores are closer together: see Figs. 4 (f) and (h), where the pores are 54 nm apart, in contrast with Figs. 4 (j) and (l), in which the pore separation is 27 nm. In other words, this is why the HCACF correlation time, $\tau$, shifts to the left in Fig. 3 (a), as the geometries become more densely packed (see corresponding geometries in Fig. 3 (d)). In short, Fig. 3(d) shows that the duration of heat flux fluctuations before the reversal process scales linearly with the distance, $d$, between the ranks of pores — indicating that the fluctuation duration depends on the time of flight to strike the pores and that pores must be causing the reversal.

Finally, a reduction in the $x$ component of the wave packets' velocity after scattering at the pore surfaces, as observed in Ref. [24] can also be observed in this work. In other words, the split velocity observed in some of the packets in Fig. 4 is likely the wave packet being scattered laterally by the curved pore, which allows for different *x*-directed velocities (i.e., in the direction of propagation explicitly shown in the heat map). Given that the simulation cell has a finite width $w$, there are only a set number of directions $\theta$ that a lattice wave of a wavelength can travel, while remaining coherent with itself across the periodic boundaries of the compute cell. Mathematically the periodic boundaries impose the condition $n\lambda = w \sin\theta$, where $n$ is an integer, $\lambda$ the wavelength of the carrier wave, and $\theta$ the direction of the wave vector relative to the long axis of the simulation cell. In Fig. 4 the wave packets have wavelengths of $\lambda =$ 13.4, 3.6, 1.5 and 1.34 nm, and the cell width is $w =$ 5.4 nm. For the longest wavelength wave packet there is no oblique path that is commensurate with the box boundaries (the equation above is only satisfied for $\theta = 0$). For the next largest wavelength of $\lambda = 3.6$ nm, there is one oblique direction possible at $\theta = 41.7°$. The wave packet traveling along this direction would have an *x* component of velocity that is 0.75 of that of the incident wave packet with $\theta = 0$, and so would leave a trace on the heat map with a slope 1.34 times steeper than the incident wave. This second possible ray is seen in Figs. 4 (f) and (j). The other waves in the wave packet would not be commensurate with the box boundaries and so the wave packet reflected along this oblique direction will be dispersed. For the wave packets with $q = 4.07$ and 5.24 oblique reflections are permitted that would leave traces on the heat map with slopes of [1.04, 1.22, 1.94] and [1.03, 1.12, 1.34, 2.18] times that of the incident wave packet, and oblique reflections corresponding to these are seen in the heat maps for these wave packets. (See Fig. S5 and the SI [40] for more details.) We note, however, that there are still some issues that are not yet clear and would be revisited in future studies, for example why the some transmissions show an increase and then a decrease as a function of $q$.

IV.   RAY TRACING MODEL

We develop a simple analytical model, as a *gedanken* experiment, to show how different types of phonon scattering manifest as signatures in the HCACF. The purpose of this is to ensure that we are correctly interpreting the anticorrelation features in the HCACF seen in panel (b) of Figs. 1–3. This model is not intended to be predictive, but to illustrate how heat trapped between the pores, as shown by the wave packet simulations (Fig. 4), can lead to the negative values in the HCACF. We consider a simple statistical model of a grey population of heat-carrying acoustic phonons that pop in and out of existence completely uncorrelated (for example from scattering with a bath of optical phonons, although the details of this are not required for this exercise) with a lifetime $\tau_o$. Each acoustic phonon contributes a stepwise heat current, $J_p(\tau)$, which has an autocorrelation function $A_p(\tau)$ (dashed blue and green lines in Fig. 5(a), respectively), that is positive and linearly decreases over time. The instantaneous heat current of the entire system is the superposition of $J_p(\tau)$ from all active phonons, but as the acoustic phonons are not correlated with one another, the system's HCACF is simply the phonon density times the average of each phonon's correlation with itself. The dashed red line in the lower panel of Fig. 5 (a) shows the integrated average of $A_p(\tau)$. To explore the effect of specular (perfectly correlated) scattering on the HCACF, we assume that each acoustic phonon experiences some scattering at a time $\alpha\tau_o$ (where $0 < \alpha < 1$) during its flight, that reflects the phonon, reversing its direction, and causing the flux and HCACF plotted with solid lines in Fig. 5 (a). This correlated scattering allows the autocorrelation to become negative. We have further considered a Poisson distribution of lifetimes $\tau_o$, and the distribution of scattering times, $P_\alpha$, such that the duration of the AC is controlled by the scattering time, $\alpha\tau_o$, and the amount of the AC is controlled through the probability of reflection at $\alpha$, $P_\alpha(\alpha)$. The Poisson distribution implies that the scattering event that annihilates the phonon is completely uncorrelated with the event that created it, and is commonly used in kinetic Monte Carlo simulations to describe the free path distribution of particles in an ideal gas [48]. Averaging over a Poisson distribution of phonon lifetimes and directions, and also the distribution of scattering times one can show that this correlated scattering reduces the thermal conductivity by

$$\frac{\kappa_x}{\kappa_o} = \int_0^1 P_\alpha (1 - 2\alpha)^2 d\alpha, \qquad (3)$$

where $P_\alpha$ is the probability distribution that a phonon is reflected at fraction $\alpha$ of the way through its flight. Additional details, and a derivation of the model are included in the Supplementary Information (SI) [40].

During diffuse scattering the incident and scattered phonon trajectories are uncorrelated which in the stochastic phonon model has the same effect as reducing the mean phonon life time $\bar{\tau}_o$. This hastens the decay of the HCACF, reducing thermal conductivity, but it does not lead to the HCACF becoming negative. Specular scattering on the other hand can be significantly more resistive for heat transport in geometries that allow anticorrelation effects, particularly if phonons live long enough after they are reflected that they undo the heat current they generated before scattering. Enforcing specularity causes anticorrelated heat flux fluctuations similar to those observed in MD.

Fig. 5 (b) shows the collective effect on correlated scattering on the net HCACF computed for a series of $P_\alpha$ distributions (plotted inset) for which the probability of scattering is shifted systematically from near the ends of the phonon flight to its middle. If the likelihood of reflection is evenly distributed throughout each phonon's lifetime ($P_\alpha = 1$, green line in Fig. 5 (b)) then the total thermal conductivity is reduced to one third of its intrinsic value, and a 10.6% dip emerges in the integrated HCACF. A larger reduction in the thermal conductivity and a more prominently peaked integrated HCACF are obtained when the probability of reflection is weighted towards the middle of the phonon flight, i.e. $\alpha = 0.5\tau_o$, to maximize the anticorrelation time (e.g., the purple and brown plots in Fig. 5 (b) show 45% and 92% reduction, respectively, similar values to those observed in some of the MD simulations). The key result from this model is that the anticorrelated heat flux observed in MD can only be achieved if phonons live long enough after they are reflected that they undo the heat current they generated before scattering. In the wave packet simulations the phonon reflections are not limited to a single back and forth oscillation, and instead multiple reflections are observed. When multiple reflections are included in the ray tracing model multiple oscillations show up in the computed ACFs. As only one dip is observed in the ACF of the Green–Kubo molecular dynamics simulations (the first panel of Figs. 1–3) this implies that, in contrast to the wave packet simulations, at 300 K phonons only remain (anti)correlated for about one reflection. One reason for this is that the wave packet simulations are performed at low temperatures and thus phonons have a much longer mean free path than in the Green–Kubo simulations at 300 K. A second probable reason is that, since the pores are cylindrical, the wave packets can reflect off the pores in different directions, as can be seen, for instance, in Fig. 4 (h). A phonon's contribution to the ACF along the $x$-direction is proportional to its velocity along $x$ squared. Thus elastic scattering that bends a wave packet away from $x$- will quash the $x$- direction HCACF.

In light of the model, we can now explain the behavior observed in Figs. 2 and 3 regarding dip 'height' and 'location'. In Fig. 2, what causes the peaks to change height is the density of inversely correlated phonons due to changes in pore size and neck width, which control the strength of the reflections. In Fig. 3, the location of the peaks shifts according to the duration for which the phonons are inversely correlated, which is in turn a function of the distance between the pores. For the geometry in green, the distance between the pores along the direction of transport, *d*, is small and, therefore, the AC effect is not visible on the HCACF. In previous work, we concluded that merely reducing the *line-of-sight* of phonons, i.e., narrowing the region available for phonon propagation, is the most important mechanism in reducing thermal conductivity in nanoporous materials [14]. The same mechanism is at play here, with the additional reduction effect due to the anticorrelation of the backscattered phonons.

Finally, we note that the effect we observe is a negatively correlated heat current which undoes its own work, which could happen in the case of coherent, or incoherent phonon propagation. Phonons need to be anticorrelated (i.e., propagate in the exact inverse direction), and in that way, they do 'interfere' with each other in that they annihilate each other in the heat transfer accountancy. This coherence is coherence over time, not the spatial coherence and wave superposition that leads to constructive/destructive interference. This is independent of wave coherent or incoherent transport conditions (superposition and constructive/destructive interference), and it can show up in either case. In fact, the simple model above demonstrates that an 'incoherent' particle phonon picture can explain this. However, as phonon transport involves a range of phonon mean-free-paths and coherence lengths, it could be possible that both effects are present. In our MD simulations it is quite possible that phonons can scatter specularly on the pores, reflect, and travel backwards to meet the previous array of pores before they undergo phonon-phonon scattering and lose coherence (the mean-free-path for scattering in Si is ~130 nm, more than twice as large as the pore separation). In that case they can interfere coherently with 'themselves' and undo their work. Physically this leads to heat trapping within the pore regions, as suggested by our wave packet simulations in Fig. 4. It is also possible that phonons with short coherent lengths can undergo an incoherent diffusive reflection, but with a sizable *x*-directed component and reflect backwards. Our results indicate that the magnitude of the AC effect is tied to the neck to pore ratio, not merely the neck size. This suggests that scattered phonons, possibly from across the spectrum, affect the negatively correlated HCACF regions. These phonons do not need to be spatially coherent, but could lose coherence after scattering and yet travel back and forth between the pore, thus giving rise to the 'AC' effect in time, essentially cancelling their contributions to thermal conductivity.

## V. CONCLUSIONS

To conclude, we have shown that special arrangements of closely packed pores in nanostructured Si can lead to anticorrelation effects in the heat flux, due to the specular backscattering of phonons at the pores. This can result in additional thermal conductivity reductions of up to ~ 80% for certain porous geometries. We surmise that AC effects could be present at experiments reporting remarkable thermal conductivity reductions in Si nanomeshes [5,49]. To investigate the origin of the anticorrelated behavior of the heat flux, we propagated wave packets through two sets of geometries (with and without AC effects). These indicate that heat can become trapped between the pores. We have observed the AC effect in (wave-based) MD simulations and have been able to replicate it with a simple (particle-based) model assuming specular reflection between the pores. Refs. [18,50] indicate that coherent reflections are only possible when surface roughness is in the order of 2–3 atomic layers, and that otherwise boundary scattering is incoherent. This is consistent with the degree of roughness in our simulations. However, while specular reflections are a necessary requirement for coherent interference [50], the mere presence of specular reflections is not in itself sufficient evidence that phonon (waves) are interfering with each other or that the AC effect is due to phonon coherence. It is possible that both coherent wave-like phonons, and incoherent particle-like phonons are present, and both to some degree undo their own work of heat propagation.

We have furthermore determined that the AC can be controlled in terms of both the amount and duration of anticorrelated specular phonon scattering. The pores provide two functions: the periodicity (along the transport direction) controls the lifetime over which a phonons' momentum is correlated, and the packing, determined by pore sizes and necks (perpendicular to transport), controls the strength of correlated phonon reflections. These functions can be engineered by tuning the spacing/periodicity between pores along the transport direction, and the pore and neck sizes, respectively. Our results suggest that the AC effect is determined by the diameter/neck ratio, and AC effects are observed for necks of at least up to ~ 6 nm. This result suggests that the porous structures can be scaled to such technologically feasible pore/neck sizes, making it easier to be used as a design tool to control thermal conductivity beyond traditional boundary scattering.

## Acknowledgements

This work has received funding from the European Research Council (ERC) under the European Union's Horizon 2020 Research and Innovation Programme (Grant Agreement No. 678763).

References:

[1] E. Pop, Nano Research **3**, 147 (2010).
[2] G. Berruto, I. Madan, Y. Murooka, G. M. Vanacore, E. Pomarico, J. Rajeswari, R. Lamb, P. Huang, A. J. Kruchkov, Y. Togawa, T. LaGrange, D. McGrouther, H. M. Rønnow, and F. Carbone, Physical Review Letters **120**, 117201 (2018).
[3] S. G. Choi, T.-J. Ha, B.-G. Yu, S. P. Jaung, O. Kwon, and H.-H. Park, Ceramics International **34**, 833 (2008).
[4] M. Lei, B. Xu, Y. Pei, H. Lu, and Y. Q. Fu, Soft Matter **12**, 106 (2016).
[5] J. Tang, H.-T. Wang, D. H. Lee, M. Fardy, Z. Huo, T. P. Russell, and P. Yang, Nano letters **10**, 4279 (2010).
[6] J. Lim, H.-T. Wang, J. Tang, S. C. Andrews, H. So, J. Lee, D. H. Lee, T. P. Russell, and P. Yang, ACS nano **10**, 124 (2015).
[7] J. A. Perez-Taborda, M. M. Rojo, J. Maiz, N. Neophytou, and M. Martin-Gonzalez, Scientific reports **6**, 32778 (2016).
[8] J. Carrete, W. Li, N. Mingo, S. Wang, and S. Curtarolo, Physical Review X **4**, 011019 (2014).
[9] R. Yang, G. Chen, and M. S. Dresselhaus, Nano Letters **5**, 1111 (2005).
[10] A. I. Boukai, Y. Bunimovich, J. Tahir-Kheli, J.-K. Yu, W. A. Goddard Iii, and J. R. Heath, in *Materials For Sustainable Energy: A Collection of Peer-Reviewed Research and Review Articles from Nature Publishing Group* (Nature Publishing Group, Singapore, 2011), pp. 116.
[11] J.-K. Yu, S. Mitrovic, D. Tham, J. Varghese, and J. R. Heath, Nature nanotechnology **5**, 718 (2010).
[12] Y. Nakamura, Science and Technology of advanced MaTerialS **19**, 31 (2018).
[13] L. Thumfart, J. Carrete, B. Vermeersch, N. Ye, T. Truglas, J. Feser, H. Groiss, N. Mingo, and A. Rastelli, Journal of Physics D: Applied Physics **51**, 014001 (2017).
[14] L. de Sousa Oliveira and N. Neophytou, Physical Review B **100**, 035409 (2019).
[15] K. D. Parrish, J. R. Abel, A. Jain, J. A. Malen, and A. J. McGaughey, Journal of Applied Physics **122**, 125101 (2017).
[16] J. Lee, W. Lee, G. Wehmeyer, S. Dhuey, D. L. Olynick, S. Cabrini, C. Dames, J. J. Urban, and P. Yang, Nature communications **8**, 14054 (2017).
[17] E. Dechaumphai and R. Chen, Journal of Applied Physics **111**, 073508 (2012).
[18] N. K. Ravichandran and A. J. Minnich, Physical Review B **89**, 205432 (2014).
[19] T. Juntunen, O. Vänskä, and I. Tittonen, Physical review letters **122**, 105901 (2019).
[20] A. McGaughey and M. Kaviany, International Journal of Heat and Mass Transfer **47**, 1783 (2004).
[21] P. Norouzzadeh, A. Nozariasbmarz, J. S. Krasinski, and D. Vashaee, Journal of Applied Physics **117**, 214303 (2015).
[22] S. Sarman and A. Laaksonen, Physical Chemistry Chemical Physics **13**, 5915 (2011).
[23] P. Keblinski, S. Phillpot, S. Choi, and J. Eastman, International journal of heat and mass transfer **45**, 855 (2002).
[24] Y. He, D. Donadio, J.-H. Lee, J. C. Grossman, and G. Galli, Acs Nano **5**, 1839 (2011).
[25] L. Maurer, S. Mei, and I. Knezevic, Physical Review B **94**, 045312 (2016).
[26] S. G. Volz and G. Chen, Physical Review B **61**, 2651 (2000).
[27] J.-H. Lee, J. Grossman, J. Reed, and G. Galli, Applied Physics Letters **91**, 223110 (2007).
[28] S. Wolf, N. Neophytou, Z. Stanojevic, and H. Kosina, Journal of electronic materials **43**, 3870 (2014).
[29] Z. Aksamija, in *ICHMT DIGITAL LIBRARY ONLINE* (Begel House Inc., 2015).

[30]     G. Romano, K. Esfarjani, D. A. Strubbe, D. Broido, and A. M. Kolpak, Physical Review B **93**, 035408 (2016).
[31]     J. R. Harter, L. de Sousa Oliveira, A. Truszkowska, T. S. Palmer, and P. A. Greaney, Journal of Heat Transfer **140**, 051301 (2018).
[32]     J. R. Harter, S. A. Hosseini, T. S. Palmer, and P. A. Greaney, International Journal of Heat and Mass Transfer **144**, 118595 (2019).
[33]     A. S. Henry and G. Chen, Journal of Computational and Theoretical Nanoscience **5**, 141 (2008).
[34]     J. Turney, E. Landry, A. McGaughey, and C. Amon, Physical Review B **79**, 064301 (2009).
[35]     S. Plimpton, Journal of computational physics **117**, 1 (1995).
[36]     F. H. Stillinger and T. A. Weber, Physical review B **31**, 5262 (1985).
[37]     P. Howell, The Journal of chemical physics **137**, 2129 (2012).
[38]     P. K. Schelling, S. R. Phillpot, and P. Keblinski, Physical Review B **65**, 144306 (2002).
[39]     J. Fang and L. Pilon, Journal of Applied Physics **110**, 064305 (2011).
[40]     See Supplemental Material at [URL to be inserted by publisher] for details on the MD simulations and the analytical model]
[41]     The negative HCACF signifies a negative correlation, between the heat fluxes — and in the strict definition of anticorrelation we have also verified that the autocorrelation of J (i.e., the HCACF) is the same as the autocorrelation of (J-<J>).
[42]     L. de Sousa Oliveira and P. A. Greaney, Physical Review E **95**, 023308 (2017).
[43]     D. Chakraborty, H. Karamitaheri, L. de Sousa Oliveira, and N. Neophytou, Computational Materials Science **180**, 109712 (2020).
[44]     L. Yang and A. J. Minnich, Scientific reports **7**, 44254 (2017).
[45]     D. Chakraborty, L. de Sousa Oliveira, and N. Neophytou, Journal of Electronic Materials **48**, 1909 (2019).
[46]     Z. Wang, J. E. Alaniz, W. Jang, J. E. Garay, and C. Dames, Nano letters **11**, 2206 (2011).
[47]     X. Wang and B. Huang, Scientific reports **4**, 6399 (2014).
[48]     A. J. McGaughey and A. Jain, Applied Physics Letters **100**, 061911 (2012).
[49]     R. Anufriev, J. Maire, and M. Nomura, Physical Review B **93**, 045411 (2016).
[50]     N. K. Ravichandran, H. Zhang, and A. J. Minnich, Physical Review X **8**, 041004 (2018).

Figure 1:

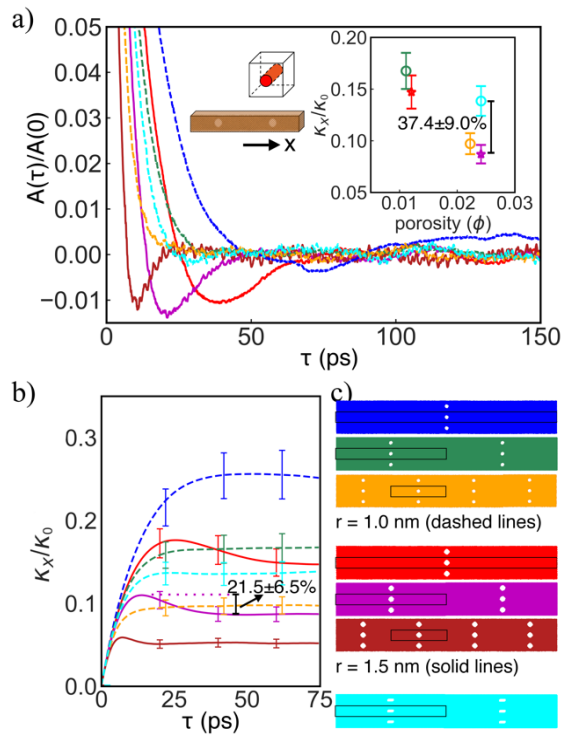

Figure 1 caption:

(a) Normalized HCACFs, $A(\tau)/A(0)$, for the geometries in c). (Inset) $\kappa_x/\kappa_0$ (extracted at the 75 ps cutoff) as a function of porosity. (b) Evolution of $\kappa_x/\kappa_0$ as a function of the HCACF time, $\tau$. The dashed/solid lines correspond to geometries without/with anticorrelated heat flux (r = 1 nm/r = 1.5 nm). The cyan geometry has an elongated pore, with a vertical (y-direction) spacing of 3.43 nm such that the 'neck' size is equivalent to the geometries with r = 1 nm. It has the same porosity and number of scatterers as the purple geometry. The error bars correspond to the standard error across the simulations performed for each geometry. (c) Cross section of the *xy*-plane (for a 108.6 nm width), for the geometries plotted in a) and b). The actual simulation cell is indicated by the black box.

Figure 2:

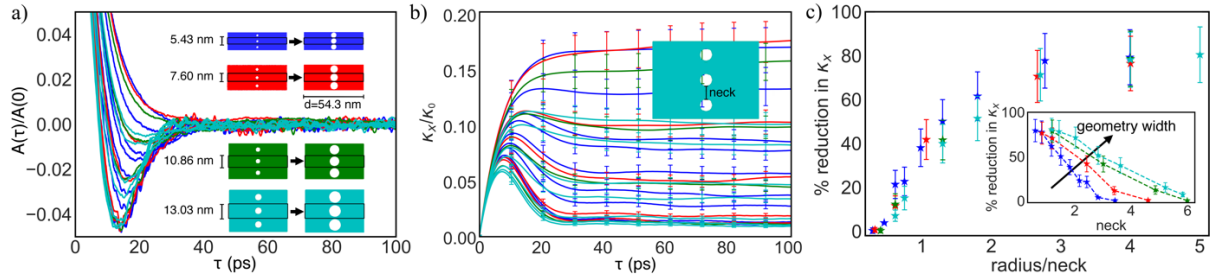

Figure 2 caption:

(a) $A(\tau)/A(0)$ for a range of geometries with the same number of pores, but varying pore radius and neck size. Simulation cell sizes are indicated in the inset. Pore radii vary between 1 and 2.5 nm for the geometries in blue, 1.5 and 3.59 nm for the geometries in red, 2.44 and 4.83 nm for the geometries in green, and 3.6 and 5.92 for the geometries in cyan. (b) Evolution of $\kappa_x/\kappa_0$ as a function of $\tau$ for the same geometries. (c) Plot of the percentage reduction in $\kappa_x$ due to the AC effect as a function of the radius to neck ratio. (Inset) Reduction in $\kappa_x$ as a function of the neck.

Figure 3:

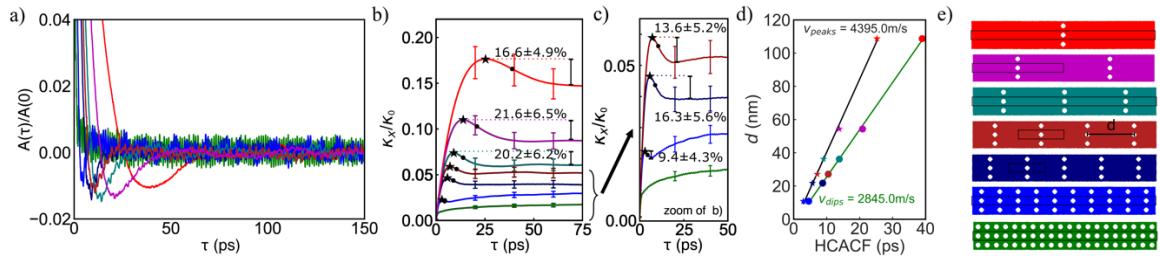

Figure 3 caption:

(a) $A(\tau)/A(0)$ for the geometries in e). All geometries have the same radius, r = 1.5 nm, but different pore concentrations. (b) Evolution of $\kappa_x/\kappa_0$ as a function of $\tau$ for the same geometries. The percentage change in $\kappa_x$ due to AC is also indicated. (c) Zoom in of the higher porosity geometries. (d) Plot of the HCACF time at which point each HCACF dip minima (circles) occurs as a function of the horizontal (x-axis) distance between the pores, and corresponding linear fit (green line); equivalent plot for the HCACF dip maxima (corresponding to where the HCACF becomes negative) (stars/black line). The slope of each line is also shown. The circles and stars match the color of the geometries. (e) Cross section of the xy-plane for the geometries plotted in a), b) and c).

Figure 4:

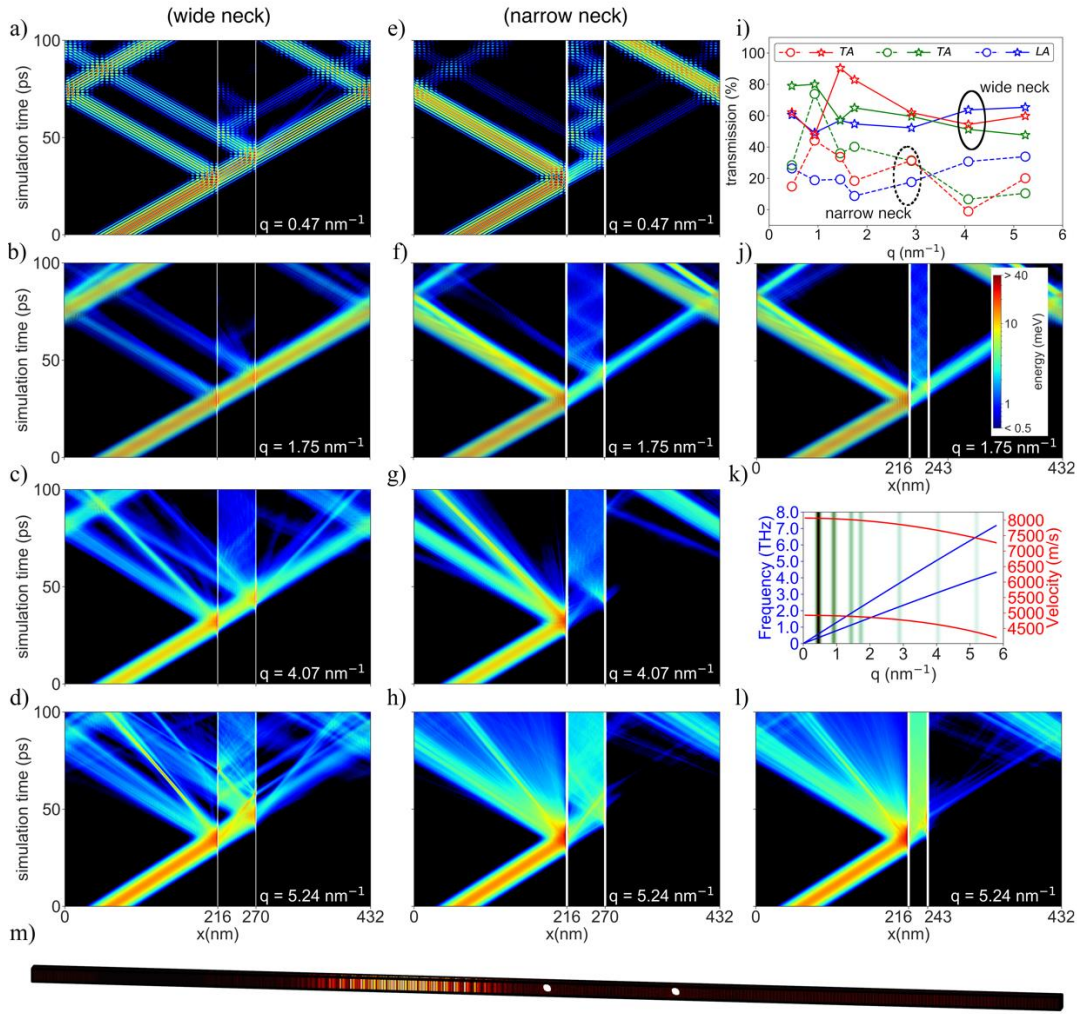

Figure 4 caption:

(a)–(h) Heatmap of the evolution of the wave packet kinetic energies during the simulation time along the width of the nanoporous geometries. The heatmap is shown for two systems with pore radii of 2 nm (left-hand-side plots (a)–(d)) and 1 nm (right-hand-side plots (e)–(h)) with corresponding 1.4 and 3.4 nm necks. The geometries are 800×10×10 unit cells and the kinetic energy is summed over each 10×10 cross-section. The white lines indicate the pore location along the width (*x*-direction) of the geometry. Each plot corresponds to a wave packet centered at a given wave vector, $q_o$, as indicated in the bottom right side of each plot for the perpendicularly polarized transverse mode (i.e., with polarization perpendicular to the height of the cylindrical pores). The distance between the pores is 54 nm. (i) Transmissions (i.e., the amount of kinetic energy that goes through) on the left-hand-side pore, for the 2 nm (dashed lines with circle markers) and 1 nm (solid lines with star markers) pore geometries obtained with the longitudinal (blue lines) and transverse modes polarized perpendicularly (indicated by the symbol ⊥, and corresponding to the red lines) and in parallel (indicated by the symbol ∥, and corresponding to the green lines). (j) Heatmap for a wave packet centered at $q_o = 1.75$ nm$^{-1}$ with a pore distance of 27 nm. This figure includes a color bar for figures (a)–(h), (j), and (l). (k) Dispersion relation showing the acoustic transverse and longitudinal mode frequencies, as well as velocity, obtained with the Stillinger–Webber potential. The vertical green shaded stripes indicate the width and weighing of each mode around $q_o$ used to construct the wave packet. (l) Heatmap for a wave packet centered at $q_o = 5.24$ nm$^{-1}$ with a pore distance of 27 nm. (m) Example of geometry with a packet propagating through it. The geometries are 800×10×10 unit cells, and the first pore is located in the middle of the supercell.

Figure 5

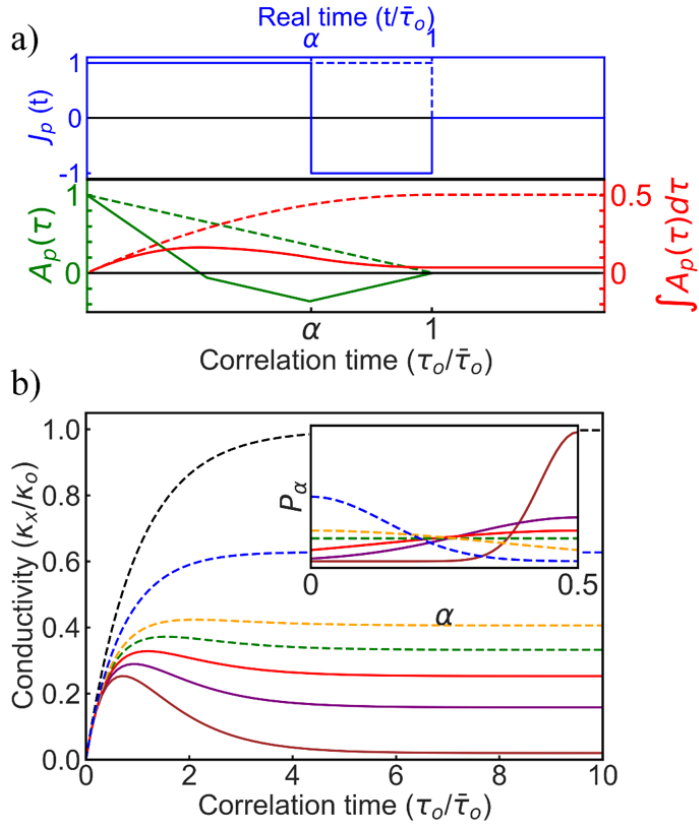

Figure 5 caption:

Stochastic model results. a) (solid blue) contribution to the heat current from a single phonon with lifetime $\tau_o$ that is reflected after time $\alpha\tau_o$, its ACF (solid green), and the integral of its ACF (solid red). The dashed lines show the corresponding functions if the phonon was not reflected. b) The net HCACF averaged over all $\tau_o$, and $\alpha$ for varying scattering probability distributions (inset). The black line is for no scattering. The green curve is for scattering with a uniform probability in $\alpha$, and shows a 10.6% dip in the ACF due to anticorrelation. In purple and brown, we show the case where the scattering probability is strongly skewed to the middle of the phonon lifetime with much larger dips in the ACF.

24